\documentclass{aa}

\usepackage{graphicx}

\begin{document}
\thesaurus{03(11.09.2;11.19.1;11.17.3;11.05.1;11.06.1)}

\title{Black hole to bulge mass correlation in Active Galactic Nuclei: A test for the simple unified formation scheme}

\author{Y. P. Wang \inst{1,2} \and P. L.\ Biermann \inst{3} \and A. Wandel \inst{4} }
\institute{Purple Mountain Observatory, Academia
   Sinica, 210008 Nanjing, China \and National Astronomical
 observatories, Chinese Academy of Sciences \and
 Max-Planck-Institut f\"ur Radioastronomie, Auf dem H\"ugel
 69, D-53121 Bonn, Germany \and Racah Institute of Physics, The Hebrew University, Jerusalem, Israel}
\offprints{ypwang@pmo.ac.cn}
\date{Received ???, Accepted ???}
 \maketitle\markboth{ Wang et al. : Black hole to bulge mass correlation}{ Wang et al. : Black hole to bulge mass correlation}

\begin{abstract}
A mass correlation of central
black holes and their spheroids $\sim 0.002$ (within a factor of three) is suggested by Hubble Space Telescope (HST) and various ground-based CCD photometries of early type galaxies. The near-IR images of quasar hosts and the emission line measurements of Broad Line Region for bright QSOs present a similar correlation, which supports the speculation of an evolutionary linkage between the early active QSO phase and the central black holes in normal galaxies. On the other hand,
recent reverberation mapping of a sample of Seyferts shows a broad distribution of black hole to bulge mass ratio with a mean of $\sim 10^{-3.5}$, about one magnitude lower than the value in early type galaxies and bright QSOs. Adopting a simple unified formation scheme for QSOs and Seyferts,
we will discuss in this letter the dependence of the black hole to bulge mass ratio in Active Galactic Nuclei (AGNs) on the
environmental parameters of the host galaxies. We show a broad distribution of the mass correlation 
could be due to different velocity dispersion of the accreting gas from different formation mechanism, and the mass ratio in normal galaxies and bright QSOs is
probably a limit case of black hole evolution by merger enhanced accretion close to Eddington limit. 

\end{abstract}

\section{Introduction}
The existence of supermassive black holes, at least in elliptical and bulge-dominated galaxies
is suggested by various observations, such as the optical spectroscopy, 
VLBI water maser measurements and X-ray observations of AGNs (Nakai et al. 1993, Kormendy \&
Richstone 1995, Faber et al. 1997, Miyoshi et al. 1995, Greenhill et al. 1996, Pounds et al. 1990). An intensive 
discussion about the
possible relationship between the central properties and their host galaxies was spurred on especially
since the high resolution HST photometry and several ground-based
CCD photometries of early type galaxies . Although there are significant scatters, a linear mass correlation of central black holes
and their host spheroids has still a high probability. 

The theoretical interpretation for such linear scale is discussed by several authors (Merritt 1998, Silk \& Rees 1998, Wang \& Biermann 1998). Considering various observations, Wang \& Biermann
(1998) proposed a possible formation scenario for elliptical and bulge-dominated galaxies, as well as
the consequential active evolution phase where 
early type galaxies are the products of major mergers between two comparable disk galaxies; the
violent collision between two galaxies could destroy the original stellar disks and form the spheroidal component of merging
galaxies after the relaxation; help to release the angular moment of the cold gas outside
and drive them inwards; a central starburst and QSO accretion in the central condensed
gas disk will compete for the gas supply, feedback and drain the gas in the disk in a short time; possibly 
grow a supermassive 
black hole during the spheroid formation, blow up the gas left probably by the nuclear wind when the central
engine gets to be powerful enough; thus 
restrict the central black hole mass and the
mass of the stellar component to a ratio of $\sim 10^{-3}$.

The numerical simulation by Wang \& Biermann (1998) shows the star formation approximately scales with the nuclear accretion
during galaxy interactions, which can regulate the black hole to bulge mass ratio to the observed level $\sim 0.002$ within a factor of three. In this scenario, a quasar black hole is 
possibly formed at a cosmic time scale of
$\sim 10^9 \rm yr$, corresponding to $z=3\sim 5$ in a flat Einstein-de Sitter world model with $\Omega=1,\, H_{0}=50\,\rm km\,s^{-1}\,Mpc^{-1}$. Afterwards, when the quasar active phase has ceased 
and the spheroidal stellar system gets relaxed in a comparable relaxation time scale,
an elliptical or spiral bulge harboring a massive black hole in the center with the mass in a factor $\sim 10^{-3}$ of the
host spheroid may appear in the universe.

We should mention the black hole evolution in this model is assumed to reach the Eddington accretion rate whenever
there is sufficient gas supplied to the center.
The question of whether this mass correlation is universal for all AGNs, i.e. whether Seyfert
galaxies follow a similar black hole to bulge mass correlation as in early type galaxies and luminous
QSOs, is not only important for the understanding of the correlation between the host galaxies and the evolution of 
active nuclei, but for the relation of the formation and evolution scheme between Seyferts and QSOs.

Recent reverberation mapping of a sample of Seyferts with reliable central masses and the bulge
magnitudes by Wandel (1999), suggests that there is a broad distribution of black hole to bulge mass ratio with a mean
of $\sim 10^{-3.5}$, about one magnitude lower than the value in early type galaxies and bright QSOs. Although this discrepancy
may in part be explained by systematic errors and selection or orientation
effects (Wandel 2000), a significant difference probably remains, in particular between Seyfert 1 galaxies and QSOs, 
at least a large dispersion for the medium-bulge
system, with masses similar to those of Seyferts (Magorrian et al. 1998, Richstone et al. 1998,
Ferrarese \& Merritt 2000). Our purpose is to firstly present a model which could explain the black hole to bulge mass correlation in
AGNs and the dependence on the environmental
parameters of the host galaxies, such as the gas or stellar velocity dispersion, as well as the relation of the central starburst and accretion process
during galactic interaction; Secondly, discuss the dispersion of black hole to bulge mass ratio in QSOs and Seyferts within a
simple unified formation scheme,
where the bulge formation and nuclear activity are triggered by galaxy mergers or tidal interactions. We found the variation of the
velocity dispersion of accreting gas could cause a range of distribution for the mass ratio, leading to a correlation of the nuclear
black hole mass and the gas velocity dispersion roughly as $M_{\rm bh}\propto v_{\rm e}^{4.7}$ from our simulation, within the slope 
suggested by recent work of Ferrarese \& Merritt (2000) and Gebhardt et al. (2000).

\section{Black hole to bulge mass correlation in AGNs} 
\subsection{Illustrative model \label{illu}}
AGNs are usually thought to be powered by gas accretion to the central massive
black holes. There is now firm evidence that QSOs could be formed during the violent galactic interactions together with central
starbursts (Sanders et al. 1988a, 1998b, Taniguchi \& Shioya 1998, Heckman et al. 1986, Bahcall et al. 1997). Although only $\sim\,10\%$
of Seyfert galaxies have companions, there are still many aspects of observations which support a simple unified formation
scheme for QSOs and Seyferts (Rafanelli et al. 1995, Taniguchi 1999). In this sense, QSOs and Seyferts are the consequences of either galaxy mergers or tidal interactions, meanwhile
a massive black hole may grow during the process of mergers or tidal interactions since the enhanced cloud-cloud collision or
star formation could increase the mass inflow to the center.

We adopt such a unified formation scheme for Seyferts where we think the tidal perturbation of some satellites not only kinematically
heat and thicken the host disks, form the central bulge, but may increase the effect of self-gravity in the molecular disk, enhance the cloud-cloud collision
and star formation, thus result in a considerable mass inflow to the central region on a relatively short time (Lin et al. 1988,
Quinn et al. 1993, Walker et al. 1996, Vel{\'a}quez \& White 1999). Although we consider in our model the tidal interactions 
as a formation scheme
for most Seyferts, we do not reject a possibility that some of AGNs with low black hole to bulge mass ratio could
still be major mergers, but at the earlier evolutionary phase in which the black holes have not had time to reach the
asymptotic value.

We adopt $\Sigma(R)=\frac{\Sigma_{0}\,R_{0}}{R}$ as 
the initial mass distribution for protogalaxies in our model, with the typical mass of $10^{11}\,M_{\odot}$ and the disk scale of 
$\sim 14\,\rm kpc$, similar as the size of our Galaxy.
The evolution of the surface density 
$\Sigma(R,t)$ of a differentially rotating disk with angular velocity $\Omega$ and 
viscosity $\nu$ is governed by (L{\"u}st 1952, Pringle 1981)
\begin{eqnarray}
\label{acc}
\frac{\partial \Sigma}{\partial t} =
-\frac{1}{R}\frac{\partial}{\partial R} \left\{
\left[ \frac{\partial (R^2
\Omega)}{\partial R} \right]^{-1}
\frac{\partial}{\partial R} \left( \nu
\Sigma R^3 \frac{\partial \Omega}{\partial R}\right) \right\}\nonumber\\
 -\frac{\Sigma}{t_{\ast}}\left(1-R_{\rm e} \right)
\end{eqnarray}

\noindent where $\nu=\beta_{1}\,v_{\phi}\,r$ for a Keplerian selfgravitating disk (Duschl et al. 2000); the accretion time scale 
$t_{\rm acc} = r^2/\nu = r/\beta_{1}v_{\rm \phi} $; and the star formation time scale $t_{\ast}=\alpha t_{\rm acc} 
= \beta_{2} r/v_{\rm\phi}$ (Pringle 1981). $\Sigma/t_{\ast}$ is the star formation rate in the disk with mass return rate
$R_{\rm e}\sim 0.3$ (Tinsley 1974). In this case, the mass influx at 
the inner boundary $R_{\rm in}$ is:

\begin{equation}
F =2\pi\beta_{\rm 1}V_{\rm 0}\left[ 2R\Sigma + R^{2}\frac{\partial
\Sigma}{\partial R}\right]_{R_{\rm in}}=2\pi\beta_{1}V_{0}\Sigma_{0}R_{0}
\end{equation}
\noindent We adopt $v_{\phi}\propto V_{0}$ for a flat rotation law for the protogalaxies in our calculation with $V_{0}=100\,\rm km/s$ as a normalization velocity. The
justification of such simplification for a self-gravitating disk at parsec to kiloparsec region is discussed in detail by 
Wang \& Biermann (1998).

The spheroidal Bondi accretion rate 
is given by $\dot{m}=4\,\pi\,\lambda\,\rho\,R_{\rm acc}^2\,v_{\rm e}$, where $R_{\rm acc}$ is the Bondi accretion radius defined as $R_{\rm acc}=\frac{G\,M_{\rm bh}}{v_{\rm e}^2}$ ($M_{\rm bh}$ is the mass of the central black
hole; $v_{\rm e}$ the effective relative velocity between seed black hole and the ambient
gas) (Bondi 1952). 
In our model, we start from a tiny black hole $\sim 5\,M_{\odot}$, with the inner boundary of the accretion disk $R_{\rm in}\sim 0.1 \rm pc$ 
which is usually thought to be the inner edge of a torus. Changing the scale of inner boundary or the seed black hole
mass only influences the early evolution, but no significant effect on the final result.
At early time, $R_{\rm in} >> R_{\rm acc}$,
we assume the mass shears inwards at the inner radius to form a uniform Bondi flow with the mass distribution $\frac{\triangle \rho(r)}
{\triangle t}=\frac{2\pi\beta_{1}V_{0}\Sigma_{0}R_{0}}{4/3\pi\,R_{\rm in}^3}$ ($\triangle t$ is the time step of the calculation); afterwards, black hole would grow and reach a stage of
$R_{\rm acc} > R_{\rm in}$, we assume in this case black hole accretes the mass shearing inwards within its influence at $R_{\rm acc}$ via a uniform Bondi flow with the mass distribution $\frac{\triangle \rho(r)}{\triangle t}=\frac{2\pi\beta_{1}V_{0}\Sigma_{0}R_{0}}{4/3\pi\,R_{\rm acc}^3}$.

The Bondi parameter is assumed to be $\lambda = 1/2$ in our calculation as a possible reduction factor due to the angular 
momentum. Thus the accretion rate $\dot{m}$ and the star formation rate $\Psi$ in a time step $\triangle t$ is given by: 
\begin{eqnarray}
\label{arate}
\dot{m}=\left\{ \begin{array}{ll}
3\,\lambda\,\beta_{1}\,V_{0}\,2\,\pi\,\Sigma_{0}\,R_{0}\,G^2\,M_{\rm bh}^2/R_{\rm in}^3\,v_{\rm e}^3 & \textrm{if $R_{\rm acc} < R_{\rm in}$}\\
\\
3\,\lambda\,\beta_{1}\,V_{0}\,2\,\pi\,\Sigma_{0}\,R_{0}\,v_{\rm e}^3/G\,M_{\rm bh} & \textrm{if $R_{\rm acc} > R_{\rm in}$} \end{array} \right.
\end{eqnarray}

\begin{eqnarray}
\Psi &=& \int_{R_{\rm in}}^{R_{out}} 2\pi\,r\,dr\,\frac{\Sigma}{t_{\star}}\nonumber\\
 &=& \frac{(1-R_{e})\,2\pi\Sigma_{0}R_{0}\,V_{0}}{
\beta_{2}}\,ln(R_{out}/R_{\rm in})
\end{eqnarray}

We could roughly estimate the black hole to bulge mass ratio by
$\frac{\dot{m}}{\Psi}=\frac{3\lambda\beta_{1}\beta_{2}v_{\rm e}^3}{GM_{\rm bh}\,ln(R_{out}/R_{\rm in})}$, where we take the value of $\dot m$ in case of $R_{\rm acc} >R_{\rm in}$. The 
estimated ratio shows a strong dependence on the velocity dispersion $v_{\rm e}$, the value $\beta_{1}*\beta_{2}=t_{\star}/t_{\rm acc}$ 
and the black hole mass $M_{\rm bh}$.

From Eq. (\ref{arate}), we know the accretion rate $\dot{m}\propto1/M_{\rm bh}$ when $R_{\rm acc} > R_{\rm in}$, which indicates that the nuclear
activity would dim slowly as the central black hole 
becomes massive enough. In this case, a critical black hole mass would be reached when
the nuclear accretion rate decreases to a level below Eddington rate $\dot{m}_{\rm edd}=2.16\,\,10^{-4}\,(M_{\rm bh}/M_{\odot})$ 
in units of $t_{0} \sim 10^4\rm yr$. 
Assuming $\dot{m}\sim
\dot{m}_{\rm edd}$, we get the critical black hole mass $\sim \, 2.6\,\,10^7\,M_{\odot}$, with $v_{\rm e}\sim 50\,\rm km/s$ and
$\beta_{1}\sim 0.05$ corresponding to the viscous diffuse time scale of $\sim 10^9 \rm yr$ (we will discuss these particular parameters later). 
Afterwards, the accretion rate would decrease quickly to be much less than Eddington rate, and the black hole will not grow significantly.
We say this is an upper limit, simply because
we use the initial surface gas density $\Sigma(R)=\frac{\Sigma_{0}R_{0}}{R}$ throughout
the estimation. Actually, the gas would be depleted during the evolution by star formation and accretion etc., it would thus take a time scale much longer than Hubble time in order to approch such a mass or maybe never could reach such a level if there is no external trigger for the
nuclear accretion, such as galaxy mergers or tidal effects.
In any sense, the black hole growth would be locked up in a level of $\sim 10^7\,M_{\odot}$
in such an evolutionary scheme. Meanwhile, we assume in our model $\beta_{1}\beta_{2}\sim 1$, since the result of numerical simulation by Wang
\& Biermann (1998) shows a strong correlation between the starburst and accretion in merging galaxies, which could regulate the black hole to 
bulge mass ratio to a level $\sim 0.002$ within a factor of three in QSOs and early type 
galaxies. In this case, if we consider Seyferts and QSOs are formed in a similar scheme, we could adopt this result as a reasonable assumption for our present model. We now use as input the critical black hole mass $\sim \, 2.6\,\,10^7\,M_{\odot}$, the assumption $\beta_{1}\,\beta_{2}\sim 1$
and $v_{\rm e}\sim 50\,\rm km/s$ to Eq. (\ref{arate}), a mass ratio about 0.001 would be approched. Choosing $v_{\rm e}\sim 50 \,\rm km/s$ is based on the observations by Zylka et al. (1990) of molecular line emission in the central region of our Galaxy with the assumption that the velocity field
of molecular clouds would be the low limit of the sound speed of hot gas (Linden et al. 1993, Garcia-Munoz et al. 1977, Reynolds 1989, 1990). We know from Eq. (\ref{arate}) and the numerical calculation,
the variation of the velocity dispersion of accreting gas, $v_{\rm e}$, would cause a broad distribution of black hole to bulge mass ratio, which is actually shown by the observations. 
The ratio could reach $\sim 0.06$ when $v_{\rm e}\sim 180 \,\rm km/s$, close to the virial velocity of the system. In this case, the 
black hole evolution 
would follow Eddington
accretion for a longer time, resulting in a QSO phase finally. We should mention here the range ($0.001\sim 0.06$) is a crude
upper limit, since we could not include the surface density evolution by star formation and accretion in this estimation. 
The practical values should be 
given by numerical simulation shown in Fig. \ref{bondi-bh}.

\subsection{Numerical results and discussion \label{num}}

To numerically solve the partial differential
Eq. (\ref{acc}), we introduce the dimensionless variables,
$r'=r/r_{\rm 0}$, $ t'=t/t_{\rm 0}$, $ \Sigma'=\Sigma/\Sigma_{\rm 0}$
with the scale $r_{\rm 0}=1 \, \rm pc$, $t_{\rm 0} \sim 10^{4}\,\rm yr$. The
general boundary conditions $lim_{r \rightarrow R_{out}}[\Sigma(r,t)]
=0$, and zero torque at the origin, $lim_{r\rightarrow R_{\rm in}}
[G(r,t)]=0$. The viscous torque $G$ is given by Pringle (1981) as  
$G(r,t)=2 \pi \nu \Sigma R^{3} \frac{\partial \Omega}{\partial R}$, with the disk's outer radius $R_{out}\sim 14\, \rm kpc$ and the
inner radius $R_{\rm in}\sim 0.1\,\rm pc$. 

This choice of boundary conditions guarantees zero viscous coupling
between the disk and the central object, allows all mass
reaching the inner boundary to flow freely inward.  

The numerical results are shown in Fig. \ref{bondi-bh}, where we model the effects of a tidal perturbation in a few kpc region by
increasing the viscous friction. In fact, the numerical simulation by Lin et al. (1988) demonstrated that a tidal disturbance
could propagate to the nucleus from a distance of a few kpc on a time scale shorter than the tidal interaction by ``swing amplication'',
which induces a rapid increase in the effective viscosity and results in an increase in the rate of mass transfer and energy dissipation. 
We adopted in the calculation a reasonable set of parameters which could match the observations quite well. The viscous diffuse
time scale is $\sim 10^9 \rm yr\,\,(\beta_{1}\sim 0.05)$ in case of a tidal perburbation according to
the numerical result of Lin et al. (1988). We consider the tidal perturbation would enhance cloud-cloud collision, thus increase both star formation
and accretion in a self-gravitation disk. According to the numerical results of Wang \& Biermann (1998) for the QSO evolution, we adopt the assumption $\beta_{1}\,\beta_{2}\sim 1$ (i.e. the star formation
approximately scales with the accretion). In this case, $\beta_{2}\sim 20$, 
corresponding to a star formation rate $\sim 10\,M_{\odot}/\rm yr$ in the central disk, which is at a reasonable level for the observed star formation rate in Seyferts. The initial seed black hole in our model is tiny, $\sim 5\,M_{\odot}$. We know from the calculation that the black hole grows by a rate $\propto M_{\rm bh}^2$ at the beginning till
$R_{\rm acc}\sim R_{\rm in}$; afterwards, the accretion rate would decrease by $1/M_{\rm bh}$ and only a mediate black hole of $\sim 10^6\,M_{\odot}$ can be formed
if there is no external trigger.
In this case, a tidal perturbation around $5\,10^9\,\rm yr$ could help to grow a massive black hole, which corresponds to $z\sim 0.8$ in an Einstein-de Sitter Universe with $\Omega=1, 
H_{0}=50\,\rm km\,s^{-1}\,Mpc^{-1}$ (where we think the tidal interactions happen frequently, also could be a peak of Seyfert activities). 

\begin{figure}
\label{bondi-bh}
\resizebox{\hsize}{!}{\includegraphics[angle=0]{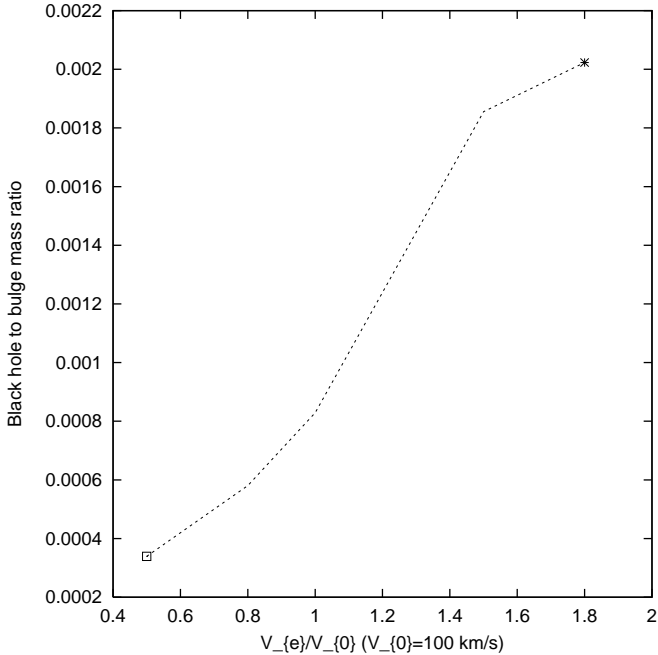}}
\caption{\label{bondi-bh} The black hole to bulge mass ratio versus velocity dispersion of accreting gas to the central region from our 
simulation. 
The square corresponds to a mass ratio at the level of the mean value of Seyfert 1 samples from reverberation mapping by Wandel (1999) and the 
velocity dispersion of $\sim 50\,\rm km/s$; The star corresponds to a mass ratio $\sim 0.002$, about the observed value in QSOs and normal 
galaxies, 
with $v_{\rm e}\sim 180 \,\rm km/s$ close to the virial velocity in the bulge system
in our calculation. We could obtain a rough proportionality of the black hole to bulge mass ratio versus the veloctiy 
dispersion of the accreting gas as $M_{\rm bh}/M_{\rm bulge} \propto v_{\rm e}^{1.4}$.}
\end{figure}

Although the parameters, such as initial black hole mass, the inner boundary of the accretion disk etc. could have
certain influence on the mass ratio, we notice from the numerical results and the illustrative estimation that velocity
dispersion of the accreting gas is
another key parameter besides the value of $\beta_{1}\beta_{2}$, which could cause a large dispersion of the black hole to bulge mass
ratio in AGNs. Fig. \ref{bondi-bh} shows a correlation between the nuclear black hole to bulge mass ratio and the velocity dispersion of 
accreting gas to the center from our simulation, which 
gives a rough proportionality of $M_{\rm bh}/M_{\rm bulge} \propto v_{\rm e}^{1.4}$. In our model, we are aiming to explain a broad distribution of black hole to
bulge mass ratio of a similar bulge system in QSOs and Seyferts. If we adopt the masses of the spheroidal systems scale with their 
luminosities as $M\propto L^{5/4}$ and the scale length $r\propto L^{1/2}$, combining with the virial theorem gives the bulge 
velocity dispersion $\sigma^2\propto \frac{M}{r}\propto 
\frac{M}{L^{1/2}}\propto \frac{M}{M^{2/5}}\propto M^{3/5}$ (Faber \& Jackson 1976, Faber et al. 1987, Peterson 1997). So, we obtain $M_{\rm bulge}\propto \sigma^{3.3}$. The assumption that the velocity dispersion of accreting gas is systematically proportional to the bulge 
velocity dispersion gives a correlation of $M_{\rm bh}\propto
v_{\rm e}^{4.7}$, which is shallower than the slope of $M_{\rm bh}\propto \sigma^{5.2}$ given by Ferrarese \& Merritt (2000), but steeper than the best-fit correlation ($M_{\rm bh}\propto \sigma^{3.75}$) by Gebhardt et al. (2000). We see from Fig. \ref{bondi-bh} that
the variation of the velocity dispersion of accreting gas would cause a broad distribution of black hole to bulge mass correlation,
where the accreting gas with higher velocity dispersion close to the bulge velocity dispersion could lead to an accretion near Eddington limit, and form more massive black holes in QSOs than in Seyferts of 
similar bulges. Therefore, the black hole to bulge mass correlation in massive systems is close to an upper limit and much tighter 
than that in medium-bulges. The physical interpretation of such discrepancy
could be due to the different formation or evolution environment of the two kinds of systems (Seyferts and QSOs). The violent collision between two galaxies could trigger intense starburst in the center, heat or shock the interstellar medium
efficiently than tidal interaction,
thus drive a Bondi flow fuelling the central black hole with probably a higher sound speed and result in a higher accretion rate. In this case, it may enhance a QSO evolution with the accretion rate close to Eddington limit, form a massive black hole, and lead to a black hole to bulge mass ratio
higher than in case of Seyferts.

\section{Summary}
Recent reverberation mapping of a sample of Seyfert 1 galaxies by Wandel (1999) suggests that the black hole to bulge mass correlation in Seyferts has a significant dispersion with a mean value of
$\sim 10^{-3.5}$, which is about one magnitude lower than in case of QSOs. Considering a simple unified formation scheme for AGNs, we demonstrated a 
possible black hole evolution scheme where we assume black hole accretes gas coming within its
influence by a uniform Bondi flow. This scenario could interpret not only the statistical mass correlation in AGNs and normal galaxies, 
but a large dispersion of the black hole to bulge mass ratio in QSOs and Seyferts. We found the black hole to bulge mass correlation in such evolution 
scheme strongly depends on the velocity dispersion of the 
accreting gas, which might be an important environmental parameter for the observed correlation besides the relation between the star
formation and accretion in the central accretion disk during galaxy interactions (Wang \& Biermann 1998, 2000). Thus, the results
of Wang \& Biermann (1998) represent a limiting case for the black hole evolution in AGN, where the accretion is close to Eddington
rate. The black hole to bulge mass
ratio versus velocity dispersion of accreting gas to the central region is plotted in Fig. \ref{bondi-bh}, which gives a rough proportionality
of $M_{\rm bh}/M_{\rm bulge}\propto v_{\rm e}^{1.4}$ for a fixed bulge system in our model. If we consider a relation of bulge mass with the galaxy velocity dispersion
in the virial equilibrium ($M_{\rm bulge}\propto \sigma^{3.3}$), we could estimate a correlation of the nuclear black hole mass with the velocity
dispersion of the accreting gas as $M_{\rm bh}\propto v_{\rm e}^{4.7}$, within the slope suggested by recent work of Ferrarese \& Merritt (2000) and  Gebhardt et al. (2000).
The square and the star in Fig. \ref{bondi-bh} correspond to the mean value of the black hole to bulge mass ratio in two typical systems (
Seyferts and QSOs), which shows the velocity dispersion of the accreting gas to form a QSO phase is much higher than in case of a Seyfert, almost close to the virial velocity of the system.
We present a possible interpretation for such scenario, where we think the intense starburst during violent mergers could heat or shock the
interstellar medium (ISM) to a higher sound speed more efficiently than by tidal interactions, 
result in a higher accretion rate close to Eddington limit and a higher black hole
to bulge mass ratio finally in QSOs and early type galaxies than in case of Seyferts.

\begin{acknowledgements}
YPW is supported by Chinese ``Chuang Xin'' project, and the Chinese post-doctor foundation. We thank our anonymous referee
for her/his helpful comments and suggestions. 

\end{acknowledgements}

\end{document}